\documentclass[prd,aps,twocolumn,tighten]{revtex4}
\usepackage{latexsym}
\begin{document}
\newcommand{\beq}{\begin{equation}}
\newcommand{\eeq}{\end{equation}}
\newcommand{\Prd}{Phys. Rev D}
\newcommand{\Prl}{Phys. Rev. Lett.}
\newcommand{\Plb}{Phys. Lett. B}
\newcommand{\Cqg}{Class. Quantum Grav.}
\newcommand{\Np}{Nuc. Phys.}

\title{Spin-2 field theory in curved spacetime}
\author{M. Novello and R. P. Neves}
\address{\mbox{}\\
Centro Brasileiro de Pesquisas F\'{\i}sicas,\\
Rua Dr.\ Xavier Sigaud 150, Urca 22290-180, Rio de Janeiro, RJ -- Brazil\\
E-mail: novello@cbpf.br}
\date{\today}

\begin{abstract}
We present a new formulation to deal with the consistency problem
of a massive spin-2 field in a curved spacetime. Using Fierz
variables to represent the spin-2 field, we show how to avoid the
arbitrariness and inconsistency that exists in the standard
formulation of spin-2 field coupled to gravity. The superiority of
the Fierz frame appears explicitly in the combined set of
equations for spin-2 field and gravity: it preserves the standard
Einstein equations of motion.\noindent
\end{abstract}

 \maketitle

\renewcommand{\thefootnote}{\arabic{footnote}}

\section{Introduction}

\subsection{General comments}

It is a rather ancient and well know difficulty of field theory the description of spin-2 field in a curved background. Although the identification of the metric structure of spacetime with gravity in Einstein general theory of relativity (GR) reduces this problem to the compatibility condition of the gravitational interaction of a spin-2 field, the question we face is more general.

The basic difficulty was pointed out by Aragone and Deser \cite{aragones71}, in the early seventies. They showed that, due to the fact that the free-field equations of motion for the spin-2 in a curved riemannian background are no longer divergence-free, as in the Minkowski case, the only acceptable situation is the one in which the background spacetime curvature is constant. This rules out compatibility with the Einstein equations, since the spin-2 field cannot be the source of a constant curvature geometry. When dealing with the possibility of non-minimal coupling, they pointed out that the Eintein equations themselves would be altered, and that severe restrictions would apply to the field variables because of the curvature's algebraic structure in each spacetime point.

In a more recent treatment of the subject, Buchbinder et al. \cite{russos} considered the case of a massive spin-2 field coupled non-minimally to gravity. They used a lagrangian written as an infinite series in the inverse mass, and showed that the set of equations thus obtained is compatible with the Einstein equations to the order of 1/m. Afterwards, they related their
findings to string theory.

In this paper we shall follow a different approach, making use of the Fierz representation of the spin-2 field \cite{fierz}. To make things clear from the start, let us remind that the spin-2 field can be described in two ways, which we will call the Einstein-frame and the Fierz-frame representations. The most common one, the Einstein-frame, uses a symmetric second order tensor $\varphi_{\mu\nu}$ to represent the field. In the Fierz-frame this
role is played by a third order tensor $F_{\alpha\mu\nu}$ which is
antisymmetric in the first pair of indices and obeys the cyclic identity. We will see that such an object must obey a further condition in order to represent only one single spin-2 field, otherwise it represents two spin-2 fields.

In the flat Minkowski spacetime both variables are equivalent: The dynamics is the same and the corresponding structure of the consistency of the dynamical equations is completely equivalent. Nevertheless, in the case of a curved spacetime this is no longer true. We shall see that the use of the Fierz-frame seems more compelling, once it yields, through the standard minimal coupling principle, a unique, non-ambiguous description. In the Einstein-frame there is an ambiguity which comes from the ordering of the
covariant derivatives, that do not commute in the curved background, as well as an ambiguity concearning the coefficients of the non-minimal coupling terms. In fact, some of the non-minimal coupling terms are exactly of the type of those that appear when one changes the order of the covariant derivatives in the equations of motion, so that different authors could see the same lagrangian as containing or not non-minimal coupling, depending of their choices of such ordering. We will show in this paper that the use of the Fierz-frame gives naturally a minimal coupling treatment for equations that are equivalent to those studied by Aragone-Deser and Buchbinder et al. as derived from a non-minimal coupling, with specific coeficients to the non-minimal coupling terms, which remain arbitrary in their formulations.

It is really a curious fact that although such ambiguity has been pointed out by many authors since its original set up, in the early fourties, the Einstein-frame has been almost uniquelly employed. As far as we know, the generalization of Fierz-frame into a curved spacetime was never used before. The reasons for this choice were related intimately with the possibility - pointed out by Gupta \cite{Gupta}, Feynmann \cite{Feynmann}, Deser \cite{Deser} and many others - of a field theoretical construction of gereral relativity, equivalent to the Einstein's geometrical approach to his gravitational theory.

The non-ambiguity of the Fierz-frame representation, in the curved spacetime background, allows us to write a set of compatible equations which are similar to those found by Buchbinder et al. \cite{russos}. Furthermore, as we use the Fierz-frame representation of the spin-2 field with minimal coupling to gravity, the Einstein equations retain their usual form. Besides, as we have alredy mentioned, our equations are equivalent to those arising from the Einstein-frame with non-minimal coupling due to the Aragone-Deser choice of derivative's ordering, with some specific coefficients - so, we are able to by-pass the problem of compatibility with the Einstein equations as pointed out by these authors \cite{aragones71}.

\subsection{Synopsis}

We start with a review of the Fierz variables to describe a spin-2 field in Minkowski spacetime. We decided to include here such a rather simple review once we realized that the majority of works on this field does not make any reference to such equivalent form of description. Then we will show in the present paper that the generalization of spin-2 field theory, from Minkowski background to arbitrary riemannian geometry, is better achieved in the Fierz
frame. Not only the theory is not ambiguous\footnote{In section III we shall describe the sort of ambiguity that appears in the Einstein-frame as it has been pointed out by previous works.} but
besides it allows a natural and complete understanding of the
necessary conditions of compatibility in a closed form. The structure of the paper is the following. After reviewing the Fierz variables in the Minkowski background in section II, we present their extension to the curved background (section III.A). We go on in section III to contrast the Einstein and Fierz representations of spin-2 field in curved spacetime (sections III.B-C), and show that the Weyl tensor plays an essencial role in the analysis of the compatibility conditions for our equations (section III.D).
We finish by making some remarks about the energy-momentum tensor for the spin-2 field in the Fierz representation (section III.E) and displaying our final set of equations (section III.F).

\section{Spin-2 field description in Minkowski background}

Since the early times of relativistic field theory it has been known that in the description of a spin-2 tensor field one can use either a second order representation that deals with a symmetric second order tensor $\varphi_{\mu\nu}$ (which we will call the Einstein representation) or by a third order tensor
$F_{\mu\nu\alpha}$ (the Fierz representation).

This second one has been forgotten, or nearly so, and thus we will make a short review of its use in the well-known case of spin-2 in Minkowski background, in order to fix notation and to show that indeed this description is as complete as the standard one. We shall argue later on that this representation is better suited for the curved background case.

\subsection{Definitions and Notation}

We define a three-index tensor $F_{\alpha\beta\mu}$ which is anti-symmetric in the first pair of indices and obeys the cyclic identity, that is

\begin{equation}
F_{\alpha\mu\nu} + F_{\mu\alpha\nu} = 0  \label{01}
\end{equation}
\begin{equation}
F_{\alpha\mu\nu} + F_{\mu\nu\alpha} + F_{\nu\alpha\mu} = 0.  \label{02}
\end{equation}

This last expression means that the dual of $F_{\alpha\mu\nu}$ is
trace-free:

\begin{equation}
\stackrel{*}{F}{}^{\alpha\mu}{}_{\mu} = 0 , \label{02bis}
\end{equation}

where the asterisk represents the dual operator defined in terms of the completelly anti-symmetric object $\eta_{\alpha\beta\mu\nu}$ by

\[
\stackrel{*}{F}{}^{\alpha\mu}{}_{\lambda} \equiv \frac{1}{2} \,
\eta^{\alpha\mu}{}_{\nu\sigma}\,F^{\nu\sigma}{}_{\lambda}.
\]

Such an object has 20 independent components. In order to eliminate the extra $10$ independent components of a three-index tensor
$F_{\alpha\beta\mu}$ that obeys conditions (\ref{01}) and
(\ref{02}) and to allow it to represent a single spin-2 field, we impose an additional requirement contained in the following lemma:

\begin{center}
Lemma 1
\end{center}

The necessary and sufficient condition that $F_{\alpha\mu\nu}$ represents an unique spin-2 field is contained in the condition\footnote{We note that this condition represents for the case of spin-2 the well-known analogous condition, for spin-1, of the existence of a potential $A_{\mu\nu}$ given by the equation
$\stackrel{*}{f}{}^{\alpha\mu}{}_{,\alpha} = 0.$}

\begin{equation}
\stackrel{*}{F}{}^{\alpha (\mu\nu)}{}_{,\alpha} = 0.  \label{03}
\end{equation}

We represent the symmetrization symbol by $A_{(\mu\nu)} \equiv A_{\mu\nu}+A_{\nu\mu}.$ Equation (\ref{03}) can be rewriten as

\begin{eqnarray}
&& {{F_{\alpha\beta}}^{\lambda}}{}_{,\mu} + {{F_{\beta\mu}}^{\lambda}}%
{}_{,\alpha} + {{F_{\mu\alpha}}^{\lambda}}{}_{,\beta} -\frac{1}{2}
\delta^{\lambda}_{\alpha} (F_{\mu ,\beta} - F_{\beta ,\mu}) + \nonumber \\
&& - \frac{1}{2} \delta^{\lambda}_{\mu} (F_{\beta ,\alpha} - F_{\alpha
,\beta}) - \frac{1}{2} \delta^{\lambda}_{\beta} (F_{\alpha ,\mu} - F_{\mu
,\alpha}) = 0.  \label{4002}
\end{eqnarray}

Just for latter use we note that a direct consequence of the above equation is the identity:

\begin{equation}
F^{\alpha\beta\mu}{}_{\, ,\mu} = 0 \ .  \label{z1}
\end{equation}

We will call a tensor that satisfies conditions (\ref{01}), (\ref{02}) and (\ref{03}) a {\bf Fierz tensor}.

If $F_{\alpha\mu\nu}$ is a Fierz tensor then it represents a unique spin-2 field. As a consequence of this lemma one can connect both representations. Condition (\ref{03}) implies that there exists a symmetric second order tensor $\varphi_{\mu\nu}$ such that we can write

\begin{eqnarray}
2\,F_{\alpha\mu\nu} &=& \varphi_{\nu\alpha,\mu} - \varphi_{\nu\mu ,\alpha} +
\left( \varphi_{,\alpha} - \varphi_{\alpha}{}^{\lambda}{}_{,\lambda}
\right)\, \eta_{\mu\nu}\nonumber \\
 &-& \left(\varphi_{,\mu} -
\varphi_{\mu}{}^{\lambda}{}_{,\lambda} \right)\, \eta_{\alpha\nu} . \label{04}
\end{eqnarray}
The factor $2$ in the lhs is introduced for convenience. The metric tensor of the Minkowski spacetime\footnote{This formula can be generalized trivially for arbitrary system of coordinates.} is represented by $\eta_{\mu\nu}.$ The proof of this Lemma is straightforward.

Taking the trace of equation (\ref{04}) we find that

\begin{equation}
F_{\alpha} = \varphi_{,\alpha} - \varphi_{\alpha}{}^{\lambda}{}_{,\lambda};
\label{05}
\end{equation}
we can thus conveniently write
\begin{equation}
2 F_{\alpha\mu\nu} = \varphi_{\nu\alpha,\mu} - \varphi_{\nu\mu ,\alpha} +
F_{\alpha}\, \eta_{\mu\nu} - F_{\mu}\, \eta_{\alpha\nu}.  \label{06}
\end{equation}

When a Fierz tensor is written under the form given in equation (\ref{04}) or (\ref{06}) we will say that it is in the Einstein frame.

\begin{center}
Lemma 2
\end{center}

A Fierz tensor $F_{\alpha \mu \nu }$ satisfies the identity

\begin{equation}
F^{\alpha }{}_{(\mu \nu ),\alpha }\equiv -\,G^{L}{}_{\mu \nu } , \label{07}
\end{equation}
where $G^{L}{}_{\mu \nu }$ is the linearized Einstein operator defined in
terms of the symmetric tensor $\varphi _{\mu \nu }$ by

\begin{equation}
G^{L}{}_{\mu \nu }\equiv \Box \,\varphi _{\mu \nu }-\varphi ^{\epsilon
}{}_{(\mu ,\nu )\,,\epsilon }+\varphi _{,\mu \nu }-\eta _{\mu \nu }\,\left(
\Box \varphi -\varphi ^{\alpha \beta }{}_{,\alpha \beta }\right) . \label{08}
\end{equation}

A manipulation of the properties of the Fierz tensor yields a direct proof of this Lemma.

Taking the divergence of $F^{\alpha }{}_{(\mu \nu ),\alpha }$ yields the identity:

\begin{equation}
F^{\alpha (\mu \nu )}{}_{,\alpha \mu }\equiv 0.  \label{07bis}
\end{equation}
Indeed,
\begin{equation}
F^{\alpha \mu \nu }{}_{,\alpha \mu }+F^{\alpha \nu \mu }{}_{,\mu \alpha }=0.
\label{070}
\end{equation}

The first term vanishes identically due to the symmetry properties of the field and the second term vanishes due to equation (\ref{z1}). As a direct consequence of this, through the use of the Lemma 2 we recover the identity which states that the linearized Einstein tensor $G^{L}{}_{\mu \nu }$ is divergence-free.

\subsection{The equations of motion}

We limit all our considerations in the present paper to a dynamics for the Fierz field which is linear. The most general linear theory is a combination of the invariants one can construct with the field. There are three of them which we represent by $A$, $B$ and $W$:

\begin{eqnarray}
A &\equiv &F_{\alpha \mu \nu }\hspace{0.5mm}F^{\alpha \mu \nu }  \nonumber \\
B &\equiv &F_{\mu }\hspace{0.5mm}F^{\mu }  \nonumber \\
W &\equiv &F_{\alpha \beta \lambda }\stackrel{\ast }{F}{}^{\alpha \beta
\lambda }=\frac{1}{2}\,F_{\alpha \beta \lambda }\hspace{0.5mm}F^{\mu \nu
\lambda }\,\eta ^{\alpha \beta }{}_{\mu \nu }  \label{AB}
\end{eqnarray}
We deal here only with the two invariants $A$ and $B$. The reason for this rests on the fact that in the linear regime the invariant $W$ does not contribute for the dynamics once it is a topological invariant, as shown in the appendix.

The standard equation for the massless spin-2 field is given by

\begin{equation}
G^{L}{}_{\mu \nu }=0.  \label{014bis}
\end{equation}
As we have seen above, in terms of the field $F^{\lambda \mu \nu}$ this equation can be written as

\begin{equation}
F^{\lambda (\mu \nu )}{}_{,\lambda }=0.  \label{014}
\end{equation}
The corresponding action takes the form

\begin{equation}
S=\frac{1}{k}\,\int {\rm d}^{4}x\,(A-B) . \label{013}
\end{equation}
Note that the Fierz tensor has dimensionality (lenght)$^{-1}$, which is compatible with the fact that Einstein constant $k$ has dimensionality (energy)$^{-1}$ (lenght)$^{-1}$. We set, from now on, $k=1.$ Then,

\begin{equation}
\delta S=\int 2F^{\alpha \mu \nu }{}_{,\alpha }\,\delta
\varphi _{\mu \nu}\,d^{4}x . \label{018}
\end{equation}
Using the identity

\begin{equation}
F^{\alpha}{}_{\mu \nu,\alpha }=\frac{1}{2}\,F^{\alpha}{}_{(\mu\nu),\alpha}=
-\,\frac{1}{2}\,G^{L}{}_{\mu \nu },  \label{01888}
\end{equation}
we obtain

\begin{equation}
\delta S=-\int G^{L}{}_{\mu \nu }\,\delta\varphi^{\mu\nu}\,d^{4}x,
\label{018bis}
\end{equation}
where $G^{L}\mbox{}_{\mu \nu }$ is given in equation (\ref{08}).

Thus, the action (\ref{013}) corresponds to

\begin{equation}
S=-\int G^{L}{}_{\mu \nu }\,\varphi ^{\mu \nu }\,d^{4}x,
\label{0181bis}
\end{equation}
in the Einstein-frame.

\subsection{Symmetry}

Let us make a remark here concerning the gauge invariance of equation (\ref{014bis}) under the map

\begin{equation}
\varphi_{\mu\nu} \rightarrow \tilde{\varphi}_{\mu\nu} = \varphi_{\mu\nu} +
\Lambda_{\mu ,\nu} + \Lambda_{\nu , \mu} . \label{30011}
\end{equation}
The field $F_{\alpha\beta\mu}$ does not change under this map only in the particular case when the vector $\Lambda_{\mu}$ is a gradient. Although the field $F_{\alpha\beta\mu}$ is not invariant under the general map when the vector $\Lambda_{\mu}$ is {\bf not} a gradient, the corresponding dynamics is invariant. Indeed, we have

\begin{equation}
\delta F_{\alpha\beta\mu} \equiv \tilde{F}_{\alpha\beta\mu} -
F_{\alpha\beta\mu} = \frac{1}{2} \, X_{\alpha\beta\mu}{}^{\,\lambda}{}_{\,
,\lambda},
\end{equation}
where

\begin{eqnarray}
X_{\alpha\beta\mu}{}^{\lambda} &\equiv&(\Lambda_{\alpha ,\beta} -
\Lambda_{\beta ,\alpha}) \delta^{\lambda}_{\mu} +
[ \Lambda^{\sigma}{}_{,\sigma} \delta^{\lambda}_{\alpha} -
\Lambda_{\alpha}{}^{,\lambda}] \eta_{\beta\mu} \nonumber \\
&-&[\Lambda^{\sigma}{}_{,\sigma} \delta^{\lambda}_{\beta} -
\Lambda_{\beta}{}^{,\lambda} ] \eta_{\alpha\mu},  \label{119}
\end{eqnarray}
and then it follows that

\begin{equation}
2\delta F_{\alpha} = X_{\alpha}{}^{\, \lambda}{}_{\, ,\lambda},
\end{equation}
whith

\[
X_{\alpha}{}^{\,\lambda} \equiv X_{\alpha\beta}{}^{\,\beta\lambda}.
\]

As a consequence of this transformation, the invariants $A$ e $B$ change accordingly:

\begin{equation}
\delta A = F^{\alpha\beta\mu}X_{\alpha\beta\mu}{}^{\,\lambda}{}_{\, ,\lambda}
\end{equation}
and

\begin{equation}
\delta B = F^{\alpha}X_{\alpha}{}^{\,\lambda}{}_{,\lambda}.
\end{equation}
We remark that $X_{\alpha\beta\mu}{}^{\lambda}$ is {\bf not} cyclic
in the indices $(\alpha\beta\mu )$, but the quantity

\begin{equation}
X_{\alpha\beta\mu}{}^{\,\lambda}{}_{,\lambda}
\end{equation}
has such cyclic property:

\begin{equation}
X_{\alpha\beta\mu}{}^{\,\lambda}{}_{,\lambda} +
X_{\beta\mu\alpha}{}^{\,\lambda}{}_{,\lambda} +
X_{\mu\alpha\beta}{}^{\lambda}{}_{,\lambda} =0.
\end{equation}
Besides, it is straightforward to show the associated identities:

\begin{equation}
X^{\alpha\beta\mu\lambda}{}_{,\lambda \alpha} = 0
\label{id}
\end{equation}
\begin{equation}
X^{\alpha\beta\mu\lambda}{}_{,\lambda \mu} = 0
\end{equation}
\begin{equation}
{X^{\alpha\lambda}{}_{,\alpha\lambda} = 0}.
\end{equation}
Thus,

\begin{equation}
\delta A = [\varphi^{\mu\alpha ,\beta} + F^{\alpha}\,\eta^{\mu\beta}] \
X_{\alpha\beta\mu}{}^{\lambda}{}_{\,,\lambda}  \nonumber,
\end{equation}
or, equivalently,

\begin{equation}
\delta A = \varphi^{\mu\alpha ,\beta} X_{\alpha\beta\mu}{}^{\,\lambda}{}_{\,
,\lambda} + F^{\alpha} X_{\alpha}{}^{\,\lambda}{}_{\, ,\lambda}.
\end{equation}
Then

\begin{equation}
\delta (A - B) = \varphi_{\mu\alpha ,\beta} X^{\alpha\beta\mu\lambda}{}_{\,
,\lambda},
\end{equation}
and

\begin{equation}
\int \varphi_{\mu\alpha ,\beta} X^{\alpha\beta\mu\lambda}{}_{\, ,\lambda} = \int
div - \int \varphi_{\mu\alpha} X^{\alpha\beta\mu\lambda}{}_{ ,\lambda \beta},
\end{equation}
so that, because of (\ref{id}),

\begin{equation}
\int \delta (A - B) = 0.
\end{equation}

This shows that the transformation

\[
F_{\alpha\beta\mu}\rightarrow F_{\alpha\beta\mu} +
X_{\alpha\beta\mu}{}^{\,\lambda}{}_{\,,\lambda}
\]
for $X_{\alpha\beta\mu}{}^{\lambda}$, given in equation (\ref{119}), leaves the dynamics invariant.

\subsection{The massive case}

Using the Fierz-frame to describe the case of massive spin-2 field in Minkowski
background, the lagrangian takes the form

\begin{equation}
L=A-B+\frac{m^{2}}{2}\,\left(\varphi _{\mu \nu }\,
\varphi ^{\mu \nu}-\varphi ^{2}\right),  \label{x055}
\end{equation}
and the equation of motion is provided by

\begin{equation}
F^{\alpha }{}_{(\mu \nu ),\alpha }+m^{2}\,\left( \varphi_{\mu\nu}-\varphi\,\eta _{\mu \nu }\right) =0  \label{mc1}
\end{equation}
or, equivalently, by

\[
G^{L}{}_{\mu \nu }-m^{2}\,\left( \varphi _{\mu \nu }-\varphi \,\eta _{\mu\nu }\right) =0.
\]
The trace of this equation gives

\begin{equation}
F^{\alpha }{}_{,\alpha }-\frac{3}{2}\,m^{2}\,\varphi =0.  \label{mc12}
\end{equation}
The divergence of equation (\ref{mc1}) yields

\begin{equation}
F_{\mu }=0.  \label{mc121}
\end{equation}
In terms of the potencial this is equivalent to

\begin{equation}
\varphi _{,\,\mu }-\varphi ^{\epsilon }{}_{\mu \,,\epsilon }=0.
\label{mcd121}
\end{equation}

Using this back to the trace equation gives $\varphi =0.$ Consequently, from
equation (\ref{mc121}) it follows

\[
\varphi ^{\mu \nu }{}_{,\nu }=0.
\]

Thus there remains only five degrees of freedom, as it should in order that $F_{\alpha\beta\mu}$ represents a pure spin-2 field.

\section{Spin-2 field description in curved riemannian background}

\subsection{The Fierz tensor in a curved spacetime}

Since the seminal paper of Aragone and Deser \cite{aragones71}
(see also for recent development and references \cite{russos}), it is a known fact the difficulty one encounters in coupling spin-2 field with gravity in the general relativity framework. In order to set the problem in a new perspective which allows us to sugest a solution, we turn our analysis to the generalization of Fierz framework into a curved spacetime.

Let us remind that a drawback of the use of Einstein-frame to generalize the equations of motion of spin-2 field into a curved spacetime is related to the ambiguity contained in such generalization. This is due to the fact that in such formulation one deals with quantities that contain second order derivatives. If the Riemann curvature tensor is not null then the order of
such derivatives is not equivalent and one has to make an arbitrary choice. The second difficulty is a dynamical one which is related to the absence of a divergence-free condition as we shall see.

In order to deal with such problems we turn our treatment to the
Fierz-frame. We start by showing that the generalization to an arbitrary riemannian curved geometry, using the standard minimal coupling principle, of condition displayed in equation (\ref{03}) to curved spacetime, that is,

\begin{equation}
\stackrel{\ast }{F}^{\alpha (\mu \nu )}{}_{;\alpha }=0  \label{40021}
\end{equation}
does not allow a representation of the field in terms of a second order symmetric tensor that acts as a potential for $F_{\alpha\beta \mu}$ (in this expression and throughout the rest of this paper, a semicolon represents the covariant derivative in the riemannian geometry). This is not a real difficulty since such condition is substituted by another one, as we will now show.

Using the minimal coupling principle let us write the field $F_{\alpha \beta\mu }$ in terms of a potential $\varphi_{\mu\nu}.$ This can be done directly from formula (\ref{06}) without any ambiguity by replacing the derivative - represented by a comma - by a covariant derivative in an arbitrary curved spacetime. We set

\begin{equation}
2F_{\alpha \beta \mu }=\varphi _{\mu \alpha ;\beta }-\varphi _{\mu \beta;\alpha }+F_{\alpha }g_{\beta \mu }-F_{\beta }g_{\alpha \mu}.  \label{x02}
\end{equation}
Thanks to the fact that the covariant derivative is not commutative, a straightforward calculation yields

\begin{equation}
\stackrel{\ast}{F}^{\alpha}{}_{(\mu\nu)\,;\alpha}=\frac{1}{2}\,\left(R_{\mu \alpha\beta}{}_{\nu}^{\ast}+R_{\beta \nu\mu }{}_{\alpha }^{\ast}\right) \,\varphi ^{\alpha \beta }.  \label{x01}
\end{equation}
Instead of eq. (\ref{40021}) this is the new condition that substitutes, in the case of curved spacetime, condition (\ref{03}) in flat space and allows the description of the Fierz tensor in terms of a potential. An important further consequence of this is that the associated quantity $W$ defined by

\[
W\equiv F_{\alpha \beta \lambda }\stackrel{\ast }{F}{}^{\alpha \beta \lambda}=\frac{1}{2}\,F_{\alpha\beta\lambda}
\hspace{0.5mm}F^{\mu \nu \lambda
}\,\eta ^{\alpha \beta }{}_{\mu \nu }
\]
is no more a topological invariant and may contribute for the dynamics of the spin-2 field, taking the form of a non minimal coupling term. In the present paper we will not consider this extra invariant. We postpone the analysis of this generalization to the future.

For completeness we note that the flat-space identity - equation (\ref{z1}) - changes to the form

\begin{equation}
F^{\alpha \mu \nu }{}_{;\nu }=-\frac{1}{2}\,\left( R^{\alpha}
{}_{\epsilon}\,\varphi ^{\mu \epsilon }
-R^{\mu }{}_{\epsilon }\,\varphi ^{\alpha\epsilon }\right) .  \label{x0101}
\end{equation}

Note that in the particular case in which the background geometry is an Einstein space, that is, $R_{\mu\nu} = \Lambda \, g_{\mu\nu}$, the right-hand side vanishes and we recover the same identity as in the flat space case.

Let us examine condition (\ref{x01}) in the particular case in which the background is of a de Sitter or Anti de Sitter geometry. The Riemannian curvature tensor can be written under the form

\begin{equation}
R_{\alpha \beta \mu \nu }=\frac{R}{12}\,g_{\alpha \beta \mu \nu }.
\label{xy1}
\end{equation}
Then it follows that

\begin{equation}
R_{\alpha}{}_{\beta}^{\ast }{}_{\mu \nu}=\frac{R}{12}\,
\eta _{\alpha \beta \mu \nu}
\label{xy2}
\end{equation}
and, consequently, equation (\ref{40021}) is valid.
This means that in the case of constant curvature in which (\ref{xy1}) is satisfied (that is, when the Weyl tensor, $W_{\alpha \beta \mu \nu },$ vanishes), the connection of the Einstein and Fierz frames are contained in the same expression (\ref{40021}) as in the flat space case (\ref{03}).

\subsection{The Fierz frame representation}

It is our purpose here to show that the difficulties one encounters in coupling spin-2 field, massive or not, with gravity represented by the geometry of spacetime in GR can become more tractable if we deal with the Fierz representation. In order to support such statement, let us first of all remember the standard argumentation that deals with the spin-2 field in Einstein frame.
The equation of motion for a massive spin-2 field is traditionaly presented in the form

\begin{equation}
G^{(1)}{}_{\mu \nu }=m^{2}\,\left( \varphi _{\mu \nu }-
\varphi \,g_{\mu \nu}\right) ,
\label{x03}
\end{equation}
in which \footnote{From now on, the simbol $\Box$ represents the curved spacetime d'Alambertian.}

\begin{equation}
G^{(1)}{}_{\mu \nu }\equiv \Box \,\varphi _{\mu \nu }-
\varphi_{\epsilon(\mu ;\nu )}{}^{;\epsilon }+\varphi _{;\mu \nu }-g_{\mu \nu}
\,\left(\Box
\varphi -\varphi ^{\alpha \beta }{}_{;\alpha \beta }\right) .
\label{08b}
\end{equation}

Let us make a small comment here concerning the ambiguity of this operator $G^{(1)}{}_{\mu \nu }.$ In the passage from the Minkowski background, in which this operator takes the form of equation (\ref{08}), to its corresponding minimal coupling version, one faces a difficulty due to the non-commutativity of the covariant derivatives. The choice made by different authors is arbitrary and has no further motivation. Instead of $G^{(1)}{}_{\mu \nu }$, one could also make another choice:

\begin{equation}
G^{(2)}{}_{\mu \nu }\equiv \Box \,\varphi _{\mu \nu }-\varphi _{\varepsilon
(\mu }{}^{;\epsilon }{}_{;\nu )}+\varphi _{;\mu \nu }-g_{\mu\nu}
\,\left(\Box \varphi -\varphi ^{\alpha \beta }
{}_{;\alpha \beta }\right) ,
\end{equation}
or any combination of both.

Nothing similar if one uses the Fierz representation. In this case, as one can see by simple inspection, there is no ambiguity. This is due to the peculiarity of the fundamental object in Fierz-frame \footnote{Note however that Aragone and Deser use a first-order treatment in terms of $\Gamma^{\alpha}_{\mu\nu}$. Although this seems similar to the Firz-frame, it doesn't use all the structural richness found in the Fierz object, $F_{\alpha\mu\nu}$.}. Besides, there is also no ambiguity in the passage from Minkowski to curved spacetime in the relation of the field $F_{\alpha \mu \nu }$ to its potential. Consequently, as we see next, there is no ambiguity in the dynamics. Equation (\ref{07}) now becomes

\begin{equation}
F^{\alpha }{}_{(\mu \nu );\alpha }\equiv -\widetilde{G}_{\mu \nu },  \label{08b3}
\end{equation}
where

\begin{equation}
\widetilde{G}_{\mu \nu }\equiv \,\frac{1}{2}\left[ G^{(1)}{}_{\mu \nu}+G^{(2)}{}_{\mu \nu }\right] .  \label{08b4}
\end{equation}

Let us come back to the Einstein frame and take the divergence of equation (\ref{x03}). The consequence of such operation is a complicated expression that has a very remarkable property: it contains only first order derivatives of the field. This expression is interpreted as the curved spacetime version of the flat space compatibility condition

\begin{equation}
\left( \varphi ^{\mu \nu }-\varphi \,\eta^{\mu \nu }\right)_{,\nu}=0.
\label{x04}
\end{equation}

This causes no difficulty. The real problem appears in the next step. In the flat space the standard procedure is to use condition (\ref{x04}) in the precedent equations and to show that one can eliminate the scalar component ${\varphi }.$ If one intends to follow this path for the curved spacetime one encounters the difficulty that the expression one obtains after this operation contains second-order derivatives and cannot be taken as a true constraint. So much for the traditional Einstein frame, let us see the modifications imposed by the use of the Fierz frame.

Using the minimal coupling principle (MCP), we keep the lagrangian constructed with the invariants $A$ and $B$\footnote{We
do not consider in the present work the dynamics containing $W.$}, that is,

\begin{equation}
L=A-B+\frac{m^{2}}{2}\,\left( \varphi _{\mu \nu }\,\varphi ^{\mu \nu}-\varphi ^{2}\right) .  \label{x05}
\end{equation}
The corresponding equation of motion is

\begin{equation}
F^{\alpha \,(\mu \nu )}{}_{;\alpha }=-m^{2}\left( \varphi ^{\mu \nu} -\varphi \,g^{\mu \nu }\right) .  \label{x06}
\end{equation}
The trace of this equation gives

\begin{equation}
F^{\alpha }{}_{;\alpha }-\frac{3}{2}\,m^{2}\,\varphi =0.  \label{x06bis}
\end{equation}

\subsection{Non-minimal coupling in the Einstein frame}

In order to compare this dynamics with the one suggested by Aragone-Deser, we use the decomposition (\ref{x02}) of $F^{\alpha \mu \nu }.$ Equation (\ref{x06}) takes the form

\begin{eqnarray}
G^{(1)}{}_{\mu \nu }+R_{\mu \epsilon \nu \lambda }\,\varphi ^{\epsilon\lambda }
&-&\frac{1}{2}\,R_{\mu \epsilon }\varphi _{\nu}
{}^{\epsilon }-\frac{1}{2}\,R_{\nu \epsilon }\varphi _{\mu}
{}^{\epsilon }\nonumber \\
&=&m^{2}\,\left( \varphi_{\mu \nu }-\varphi \,
g_{\mu \nu }\right) ,  \label{x033}
\end{eqnarray}
or, equivalently,

\begin{equation}
\widetilde{G}_{\mu \nu }=m^{2}\,\left( \varphi _{\mu \nu }-\varphi \,g_{\mu\nu }\right) .
\end{equation}

First of all we note that, from the Aragone-Deser point of view, in the Fierz representation an effective non-minimal coupling between the spin-2 field and gravity appears naturally through the potential $\varphi _{\mu \nu}$.

The expression (\ref{x033}) can also be obtained in the Einstein-frame if one adds arbitrarily to the standard lagrangian definite extra terms, that is

\begin{eqnarray}
L^{(1)}=&-& G^{(1)}{}_{\mu \nu }\,\varphi ^{\mu \nu }
-\alpha\,R_{\mu \epsilon \nu \lambda }\,\varphi ^{\epsilon\lambda}
\,\varphi ^{\mu\nu }+ \beta \,R_{\mu \nu }\,\varphi^{\mu
\epsilon}\,\varphi ^{\nu}{}_{\epsilon } \nonumber \\
&+&\frac{m^{2}}{2}\,\left( \varphi _{\mu \nu }\,\varphi ^{\mu \nu}-\varphi ^{2}\right)
 \label{x0300}
\end{eqnarray}
with the choice $\alpha =\beta =1.$ Let us make two comments on this apparent equivalence. First we note that in the Einstein-frame, represented by expression (\ref{x0300}), one deals with a specific combination of the extra non-minimal curvature-dependent terms, chosen arbitrarily among all other possibilities. Contrarily to this, the Fierz-frame yields a peculiar combination of such non-minimal terms, without ambiguity. The non minimal coupling itself, actually, turns out to be a consequence of the choice of the order of the derivatives; making the choice $\alpha =\beta =1,$ this lagrangian can be written as

\begin{equation}
L^{(1)}=-\widetilde{G}_{\mu \nu }\,\varphi ^{\mu \nu }
+\frac{m^{2}}{2}\,\left( \varphi _{\mu \nu }\,\varphi ^{\mu \nu}-\varphi ^{2}\right) ,
\end{equation}
so that we have minimal coupling.\

Besides there is a further property which implies a crucial distinction between the two modes.

Assuming Einstein representation, equation (\ref{x0300}) must be implemented by the free part of General Relativity, that is

\begin{equation}
L=\frac{1}{2}\,R.  \label{x0300bis}
\end{equation}
As a consequence of this, independent variation of the metric $g_{\mu \nu }$ and the field in the Einstein-frame yield a modified equation for the metric, once the non minimal terms must be varied yielding further terms in the GR set of equations, containing derivatives of order higher than two. In the Fierz-frame, such modification does not exist and the equation of GR remains in its standard form,

\begin{equation}
R_{\mu\nu} - \frac{1}{2} \, R \, g_{\mu\nu} = -\, T_{\mu\nu}.  \label{z12}
\end{equation}
This shows the superiority of the Fierz frame: it preserves the standard Einstein equation of motion.

An important further property concerns the fact that the trace of the new terms that appear containing curvature in equation (\ref{x033}) vanishes identically. In the Einstein frame this property occurs only in the very special case $\alpha=\beta=1,$ which is precisely the combination needed to arrive at the equivalence of both representations. Note that equations (\ref {x06bis}) and (\ref{mc12}) are very similar. The importance of this fact is that equation (\ref{x06bis}) is crutial to the solution of the compatibility conditions, as we will show in the next section.

\subsection{Compatibility conditions}

The absence of a divergence-free identity in the case of a spin-2 field in arbitrary curved spacetime provokes the compatibility condition that may destroy the pure spin-2 structure of the field. At this point it has been a traditional usage the limitation of the curved background by imposing the condition that the geometry is such that the curvature is constant, that is, one imposes

\[
R_{\mu\nu} = \frac{1}{4} \, R g_{\mu\nu}.
\]

In other words, one deals with an Einstein space. This solves one part of the problem but leads to a rather dramatic situation: one cannot deal with a closed system restricted to spin-2 field plus gravity satisfying Einstein equation. This is a direct consequence of the fact that the energy-momentum tensor of the spin-2 field cannot be taken as the source of the (constant) curvature of spacetime. One could be prepared to restrict in some way the properties of the geometry which is generated by an arbitrary spin-2 field. But certainly this condition should be more soft. Take for instance the example of the geometry generated by a spin-1 field that satisfies linear Maxwell equation. It has been shown by Rainich and latter on by Wheeler \cite{wheeler} that in order of a geometry to have a spin-1 as its source it must satisfy

\begin{equation}
R_{\mu\alpha}R^{\alpha\nu}=\lambda \delta_{\mu}{}^{\nu}.
\label{qwert}
\end{equation}
One should expect to arrive at a corresponding condition for the case of spin-2, with some different kind of restriction.

Taking the covariant divergence of equation (\ref{x06}) we find

\begin{equation}
Z^{\mu }+m^{2}\,F^{\mu }=0,  \label{x07}
\end{equation}
where we define

\begin{equation}
Z^{\mu }\equiv -F^{\alpha \,(\mu \nu )}{}_{;\alpha \nu },
\end{equation}
so that

\begin{equation}
Z^{\mu }=R^{\alpha \beta \lambda \mu }\,F_{\alpha \beta \lambda }-\frac{1}{2}\,\left( R_{\varepsilon}
{}^{[\lambda}\varphi^{\mu]\varepsilon}\right) _{;\lambda }.  \label{x08}
\end{equation}

Equation (\ref{x07}) contains - as in the case of the Einstein frame - derivatives of the first order on the spin-2 field $\varphi _{\mu \nu }$. We can understand this equation as a compatibility condition in the same manner as the divergence equation in flat space. It remains to obtain a further requirement in order to eliminate the scalar freedom. The natural way should be to use this equation (\ref{x07}) into the trace of the equation of motion (\ref{x06}). However, in the curved spacetime the structure of the constraint is such that does not allow the solution of this problem from such substitution. We then turn to another procedure to obtain a scalar constraint. Let us do this by taking a second divergence of the equation of motion. A direct calculation gives

\begin{equation}
F^{\alpha \,(\mu \nu )}{}_{;\alpha ;\mu ;\nu }=\left( R^{\alpha \beta\lambda \mu }\,F_{\alpha \beta \lambda }\right) _{;\mu },  \label{x09}
\end{equation}
and hence

\begin{equation}
\left( R^{\alpha \beta \lambda \mu }\,F_{\alpha \beta \lambda }\right)_{;\mu }=-m^{2}\,F^{\alpha }{}_{;\alpha }.  \label{x09bis}
\end{equation}

If the geometry generated by the spin-2 field, through the Einstein equation, is such that the Riemann curvature tensor satisfies the condition

\begin{equation}
R^{\alpha \beta \lambda \mu }\,F_{\alpha \beta \lambda }=0,  \label{x10}
\end{equation}
then it follows that

\begin{equation}
F^{\alpha }{}_{;\alpha }=0.  \label{x11}
\end{equation}
Using this result back to the trace of the equation of motion (\ref{x06bis}) we obtain the vanishing of the scalar $\varphi $ as in the flat space case, providing a consistent theory of massive spin-2 field coupled to gravity.

Let us write equation (\ref{x10}) as

\begin{equation}
W^{\alpha \beta \lambda \mu }\,F_{\alpha \beta \lambda }+
T^{\alpha \beta}F^{\nu }{}_{\alpha \beta }+T^{\alpha \nu}F_{\alpha}-\frac{2}{3}T\,F^{\nu}=0,  \label{x12}
\end{equation}
where we made use of equation (\ref{z12}). It is quite clear from this expression that the appearence of the Weyl tensor in it is the fundamental point that makes it compatible with the Einstein equations. In the Aragone-Deser equation, (\ref{x03}), the corresponding condition involves only the Ricci tensor $R_{\mu\nu}$, which is already compromised with the Einstein equations - that is the source of the incompatibility that they pointed out. In our equation, the presence of the Weyl tensor turns it possible for the geometry to satisfy the condition stated above, without getting into contradiction with the Einstein equations.

\subsection{The energy-momentum tensor for the spin-2 field}

We showed in section III.C that, in the Fierz representation of the spin-2 field with minimal coupling to gravity, the Einstein equations remain valid. The quantity that appears at the r.h.s. of equation (\ref{z12}) is the energy-momentum tensor for the spin-2 field, given by the standard procedure:

\begin{equation}
T_{\mu\nu}=\frac{2}{\sqrt{-g}}\left\{ \frac{\delta(\sqrt{-g} L)}{\delta
g^{\mu\nu}}- \left[\frac{\delta(\sqrt{-g}L)}
{\delta g^{\mu\nu}{}_{,\varepsilon}}
\right]_{,\varepsilon}\right\},  \label{n300}
\end{equation}
where the $L$ is the lagrangian used in section
III.B. A rather long but straightforward calculation allows us to
write the energy-momentum tensor as provided by:

\begin{eqnarray}
T^{\mu \nu }= &-&L\,g^{\mu \nu }+2[2F^{\mu \alpha \beta }F^{\nu }
{}_{\alpha \beta }+F^{\alpha \beta \mu }F_{\alpha \beta }{}^{\nu
}\nonumber \\
&-&F^{\alpha (\mu \nu)}F_{\alpha }
-F^{\mu }F^{\nu }]  + [F^{\alpha (\mu \nu )}\varphi _{\alpha
}{}^{\varepsilon }\nonumber \\
&-&F^{\alpha(\varepsilon \mu )}\varphi _{\alpha }{}^{\nu
}-F^{\alpha (\varepsilon \nu)}\varphi _{\alpha }{}^{\mu
}]_{;\varepsilon }\nonumber\\
&+&2m^{2}(\varphi^{\mu\alpha}\varphi_{\alpha}{}^{\nu}-\varphi\,
\varphi^{\mu\nu}). \label{n500}
\end{eqnarray}

Note that we cannot, even for the massless case, write for $T^{\mu\nu}$ an expression containing only the tensor $F_{\alpha\mu\nu}$, because the potencial $\varphi_{\mu\nu}$ appears explicitly in the expression above (which is also true for the condition (\ref{x01})).

It is a remarkable fact that, since our equations could be equally derived in the Einstein-frame with non-minimal coupling, as we showed in section III.C (through the choice $\alpha =\beta =1$), this $T^{\mu \nu }$ contains all the terms that would alter the Einstein equations due to the non-minimal coupling in this representation.

\subsection{The final set of equations}

Thus, the complete set of consistent equations for the massive spin-2 field, in the Fierz representation, is

\begin{equation}
\left\{
\begin{array}{c}
F^{\alpha \,(\mu \nu )}{}_{;\alpha }=-m^{2}\left( \varphi ^{\mu \nu}-\varphi \,g^{\mu \nu }\right) \\
\\
R_{\mu \nu }-\frac{1}{2}\,R\,g_{\mu \nu }=-\,T_{\mu \nu } \\
\\
Z^{\mu }+m^{2}\,F^{\mu }=0 \\
\\
R^{\alpha \beta \lambda \mu }\,F_{\alpha \beta \lambda }=0,
\end{array}
\right.  \label{ww}
\end{equation}
where $T^{\mu \nu }$ is given by (\ref{n500}) and $Z^{\mu }$  by (\ref{x08}). The tensor $\varphi _{\mu \nu }$ satisfies also the condition $\varphi =0,$ which is implied by these equations.

\section{Conclusion}

We showed in this work that the generalization of the Fierz
representation of the spin-2 field to a curved spacetime has some
interesting properties: first of all, it is completely
non-ambiguous with regard to the order of the derivatives. Second,
the assumption of minimal coupling, in this representation, result
in equations that are equivalent to the equations with non-minimal
coupling in the Einstein representation, when the traditional
Aragone-Deser choice of the order of the derivatives is used, with
some fixed coefficients for the extra terms. This allows us to
solve the compatibility conditions for the massive field. At the
same time, in the Fierz representation, the Einstein equations
remain valid, with their usual form. These results indicate that
the Fierz representation seems to be the most natural way to deal
with the coupling of the spin-2 field to gravity.

\section{Appendix}

$\int \,W\, d^{4}x$ is a topological invariant. Indeed,

\begin{eqnarray}
&& Q = \int \stackrel{*}{F}{}_{\mu\alpha\beta} F^{\mu\alpha\beta} d^4V
\nonumber \\
&&= \int F^{\alpha\beta\mu}\,[\eta_{\alpha\beta}{}^{\,\rho\sigma}
\,\varphi_{\mu\rho ,\sigma} + F^{\rho}\, \eta_{\alpha\beta\rho\mu}]\,dV.
\end{eqnarray}

Since

\[F^{\alpha\beta\mu}\,\eta_{\alpha\beta\rho\mu} = 0, \]
then

\begin{equation}
Q = \int [F^{\alpha\beta\mu}\,{\eta_{\alpha\beta}}^{\rho\sigma}
\,\varphi_{\mu\rho}]_{,\sigma} - \int F^{\alpha\beta\mu,\sigma}\,
\eta_{\alpha\beta\rho\sigma}\,\varphi_{\mu\rho}.
\end{equation}

By equation (\ref{40021}) the second term vanishes and it remains:

\begin{equation}
Q = \int \partial_{\sigma} (\eta^{\alpha\beta\rho\sigma} F_{\alpha\beta\mu} {\varphi^{\mu}}_{\rho}).
\end{equation}
Or, in other way round,

\begin{equation}
W = F^{\mu\alpha\beta}\, \stackrel{*}{F}{}_{\mu\alpha\beta}
\end{equation}
can be rewriten as

\begin{equation}
W = \left(\stackrel{*}{F}{}_{\rho\sigma\beta}
\,\varphi^{\beta\rho}\right)_{,\epsilon}\, \eta^{\sigma\epsilon}.
\end{equation}

Calling

\begin{equation}
K_{\sigma} \equiv -\,\stackrel{*}{F}{}_{\sigma\alpha\beta}
\,\varphi^{\alpha\beta}
\end{equation}
we write

\begin{equation}
W = {K^{\mu}}_{,\mu}.
\end{equation}

\section{Ackowledgement}

This work was partially supported by the Brazilian agency Conselho Nacional
de Desenvolvimento Cient\'{\i}fico e Tecnol\'ogico (CNPq).

\end{document}